\documentclass[12pt]{article}

\usepackage{epsf,a4wide}

\newcommand{\eqpt}{\hspace{6pt}.\hspace{6pt}}  
\newcommand{\eqcm}{\hspace{6pt},\hspace{6pt}}
\newcommand{\GeV}{\mbox{\ GeV}}
\newcommand{\re}{\mbox{Re}}                 
\newcommand{\im}{\mbox{Im}}

\newcommand{\ta}{\theta}           
\newcommand{\veps}{\varepsilon}

\newcommand{\xp}{x_{I\!\!P}}       
\newcommand{\bh}{\hat{\beta}}

\newcommand{\lsim}{\raisebox{-.5ex}{\footnotesize$
     \,\:\stackrel{\textstyle<}{\sim}\,\:$}}

\begin{document}


\begin{flushright}
  CPT--S556--0997 \\[\baselineskip]
\end{flushright}

\begin{center}
  {\Large PHOTON POLARISATION IN DIFFRACTIVE \\[0.3\baselineskip]
    DEEP INELASTIC SCATTERING} \\[1.5\baselineskip]
  M. Diehl \\[\baselineskip]
  \textit{CPT, Ecole Polytechnique, 91128 Palaiseau, France}
  \\[1.5\baselineskip]
  \textbf{Abstract} \\[\baselineskip]
  \parbox{0.9\textwidth}{The distribution of a suitably defined
    azimuthal angle in diffractive deep inelastic scattering contains
    information on the polarisation of the exchanged photon. In
    particular it allows one to constrain the longitudinal diffractive
    structure function. We investigate the potential of such bounds in
    general and for particular diffractive final states.}
\end{center}

\vspace{0.5\baselineskip}

\section{Introduction}
The inclusive cross section for deep inelastic diffraction measured at
HERA~\cite{HERA} shows a remarkable pattern of scaling violation: the
diffractive structure function $F^{D(3)}_2(\xp,\beta,Q^2)$ is found to
rise with $Q^2$ even at rather large values of the scaling variable
$\beta$. When $F^D_2$ is interpreted in terms of diffractive parton
densities evolving according to the DGLAP equations this leads to a
significant amount of gluons with a large momentum fraction. An
important question for the QCD analysis of $F^D_2$ and also for its
extraction from the data is how much of $F^D_2$ is due to
longitudinally polarised photons. Several models of diffraction find
in fact a considerable longitudinal contribution $F^D_L$ to $F^D_2 =
F^D_T + F^D_L$ at large $\beta$~\cite{Brisk}. It is of course crucial
to know whether or not such a contribution is of leading twist if one
wants to describe $F^D_2$ in terms of leading twist parton densities
and their evolution~\cite{BarTwist}.

In~\cite{HDCB,MD} its was pointed out that an appropriate azimuthal
distribution in the final state can be used to obtain bounds on
$F^D_L$, without requiring measurements at different energies of the
$ep$ collision as in the standard method for the separation of
longitudinal and transverse structure functions. The aim of this paper
is to make some comments on the potential of these bounds in general,
and to see what can be expected for $F^D_L$ and its bounds for
particular diffractive final states and dynamical models.

\section{The azimuthal angle}

\begin{figure}[t]
  \begin{center}  \leavevmode
    \setlength{\unitlength}{1cm}
    \begin{picture}(10,3.6)
      \put(0,0){\epsfysize 3.6cm \epsfbox{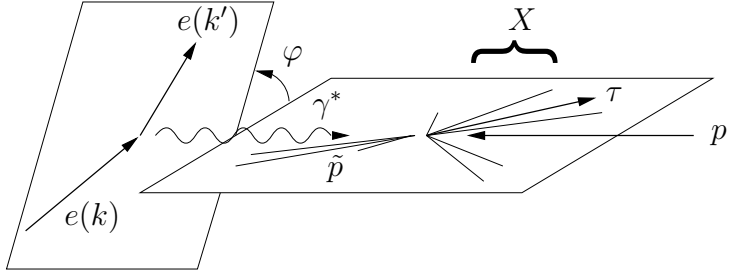}}
      \put(3.7,2.8){$\varphi$}
      \put(0.8,0.6){$e(k)$}
      \put(2.3,3.2){$e(k')$}
      \put(4.1,2.1){$\gamma^\ast$}
      \put(4.3,1.25){$\tilde{p}$}
      \put(9.4,1.75){$p$}
      \put(8.0,2.25){$\tau$}
      \put(6.7,3.2){$X$}
    \end{picture}
  \end{center}
  \caption{\label{fig:kin}Kinematics of a diffractive process in the
    $\gamma^\ast p$ c.m. The vector $\tau$ is defined
    in~Fig.~\protect\ref{fig:tau}.}
\end{figure}

\begin{figure}[b]
  \begin{center}  \leavevmode
    \setlength{\unitlength}{1cm}
    \begin{picture}(11.8,3.5)
      \put(0,0){\epsfxsize 11.8cm \epsfbox{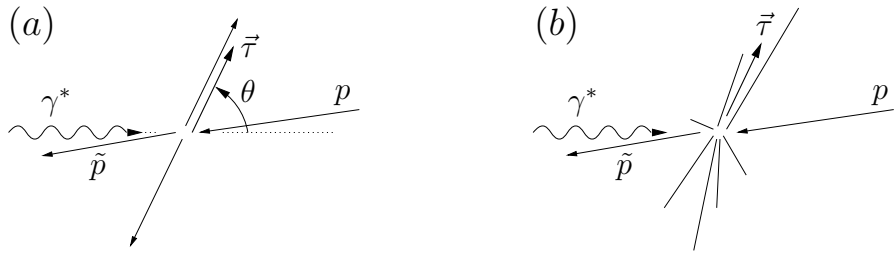}}
      \put(0,2.9){\large $(a)$}
      \put(0.45,1.9){$\gamma^\ast$}
      \put(4.35,2.05){$p$}
      \put(1.1,1.05){$\tilde{p}$}
      \put(3.1,2.6){$\vec{\tau}$}
      \put(3.1,2.0){$\ta$}
      \put(7.0,2.9){\large $(b)$}
      \put(7.45,1.9){$\gamma^\ast$}
      \put(11.5,2.05){$p$}
      \put(8.1,1.05){$\tilde{p}$}
      \put(9.95,2.9){$\vec{\tau}$}
    \end{picture}
  \end{center}
  \caption{\label{fig:tau}Definition of  $\vec{\tau}$ as the thrust
    axis in the c.m.\ of the diffractive system $X$, oriented to point
    into the photon direction. $\ta$ is the angle between $\vec{\tau}$
    and the photon momentum.}
\end{figure}

Let us consider a diffractive reaction $e(k) + p(p) \to e(k') + X(p_X)
+ \tilde{p}(\tilde{p})$, where $X$ is the diffractive system and
$\tilde{p}$ the scattered proton or proton remnant and where we have
indicated four-momenta in parentheses. We will always work in the
one-photon exchange approximation. If we define some four-vector
$\tau$ in the final state and go to the $\gamma^\ast p$ c.m.\ with the
positive $z$ axis defined by the photon momentum $q$ then we have an
azimuthal angle $\varphi$ between the electron momentum $k$ and $\tau$
(Fig.~\ref{fig:kin}) which contains information on the polarisation of
the exchanged photon. For $\tau$ we have the freedom of choice under
the condition~\cite{HDCB} that it should only depend on momenta of the
subreaction $\gamma^\ast(q) + p(p) \to X(p_X) + \tilde{p}(\tilde{p})$.
Here we choose the following: go to the rest frame of the system $X$
and set $\tau = (0, \vec{\tau})$ where $\vec{\tau}$ is the thrust axis
of $X$ oriented to point into the photon direction. If $X$ consists
only of two particles then $\vec{\tau}$ simply is the direction of the
forward particle as shown in Fig.~\ref{fig:tau} $(a)$, the general
case is represented in Fig.~\ref{fig:tau} $(b)$.

The dependence of the $ep$ cross section on this angle is explicitly
given as a trigonometric polynomial~\cite{HDCB,MD}
\begin{eqnarray}
  \label{master}
  \frac{d \sigma(ep \to e \tilde{p} X)}{d\varphi \, d Q^2 \,
      d x \, d \xp \, d\Phi} &=& \frac{\alpha_{\it em}}{2 \pi^2} \,
      \frac{1 - x}{x Q^2} \left( 1 - y + y^2 /2 \right)  \cdot \big\{
      S_{++} + \veps S_{00} - \veps S_{+-} \cdot \cos 2\varphi
  \hspace{5em} \\ \nonumber
  &&  {} - 2 \sqrt{\veps (1 + \veps)} \, \re S_{+0} \cdot \cos\varphi 
      + 2 r_L  \sqrt{\veps (1 - \veps)} \, \im S_{+0} \cdot
      \sin\varphi \big\}  \eqcm
\end{eqnarray}
where $r_L = \pm 1$ is the helicity of the incident lepton. We have
used the conventional variables $Q^2 = - q^2$, $x = Q^2 / (2 q \cdot
p)$, $y = (q \cdot p) /(k \cdot p)$, $\beta = Q^2 / (2 q \cdot
\Delta)$, $\xp = (q \cdot \Delta) /(q \cdot p)$ with $\Delta = p -
\tilde{p}$, and the usual ratio $\veps = (1 - y) /(1 - y + y^2 /2)$ of
longitudinal and transverse photon flux. The functions
\begin{equation}
  \label{S}
  S_{m n}(\xp, \beta, Q^2, \Phi) = \frac{d \sigma_{m n}}{d \xp \,
  d\Phi}   \eqcm \hspace{2em} m, n = -,0,+
\end{equation}
do \emph{not} depend on $\varphi$, for $m = n$ they are the
differential $\gamma^\ast p$ cross sections for photon helicity $m$,
and for $m \neq n$ they give the interference between photon
helicities $m$ and $n$. With $\Phi$ we have denoted any additional
variables of the $\gamma^\ast p \to X \tilde{p}$ reaction one may want
to consider, provided that they are invariant under a parity
transformation, which excludes e.g.\ further azimuthal angles.
$S_{+-}$ is real under these circumstances whereas $S_{+0}$ may have
an imaginary part, which one can however expect to be small compared
with its real part~\cite{MD}. Note that for the appearance of $S_{+0}$
in the $ep$ cross section it is essential that $\varphi$ is a genuine
azimuthal angle ranging from $0$ to $2 \pi$; if one were to define
$\varphi$ as the angle between the two planes shown in
Fig.~\ref{fig:kin} then $\varphi$ and $\varphi + \pi$ would be
equivalent and terms with $\cos\varphi$ and $\sin\varphi$ would
average out in (\ref{master}).

\section{Bounds on the longitudinal cross section}

From the $\varphi$ dependence of the $ep$ cross section one obtains
the interference terms $S_{+-}$ and $S_{+0}$ in addition to the
weighted sum $S_{\veps} = S_{++} + \veps S_{00}$ of $\gamma^\ast p$
cross sections.  These allow to constrain $S_{00}$ as~\cite{MD}
\begin{eqnarray}
&&   \label{bound-1}
  \frac{S_\veps - S_{+-}}{2 \veps} - 
  \sqrt{ \left( \frac{S_\veps - S_{+-}}{2 \veps} \right)^2 - \frac{2
  \, |S_{+0}|^2}{\veps} } \; \le
  \; S_{00}   \eqcm  \\
&&   \label{bound-2}
  S_{00} \; \le \; \frac{S_\veps - S_{+-}}{2 \veps} +
  \sqrt{ \left( \frac{S_\veps - S_{+-}}{2 \veps} \right)^2 - \frac{2
  \, |S_{+0}|^2}{\veps} }  \eqcm  \\
&&   \label{bound-3}
  S_{00} \; \le \; \frac{S_\veps + S_{+-}}{\veps}  \eqpt
\end{eqnarray}
If $\im S_{+0}$ is unknown because the lepton beam is unpolarised one
can replace $S_{+0}$ with $\re S_{+0}$ here and in the sequel. Weaker
but simpler versions of bounds (\ref{bound-1}) and (\ref{bound-2}) are
\begin{eqnarray}
&&  \label{bound-4}
  \frac{2 \, |S_{+0}|^2}{S_\veps - S_{+-}} \; \le \; S_{00}  \eqcm \\
&&  \label{bound-5}
  S_{00} \; \le \; \frac{S_\veps - S_{+-}}{\veps}  \eqcm
\end{eqnarray}
respectively, they correspond to the leading terms when
(\ref{bound-1}) and (\ref{bound-2}) are Taylor expanded in $| S_{+0}
|^2 /(S_\veps - S_{+-})^2$.


To obtain bounds on the $\Phi$-integrated longitudinal cross section
$d\sigma_{00} /d \xp$, from which the diffractive structure function
$F^{D(3)}_L(\xp, \beta, Q^2)$ is obtained by multiplying with $Q^2
\cdot \linebreak[0] (1 - x) /(4 \pi^2 \alpha_{\it em})$, there are two
possibilities:
\begin{enumerate}
\item one can extract $S_{\veps}$, $S_{+-}$, $S_{+0}$ differential in
  certain variables $\Phi$, evaluate the bounds (\ref{bound-1})
  to~(\ref{bound-5}) on $S_{00}$ and then integrate over $\Phi$, or
\item one can determine $S_{\veps}$, $S_{+-}$, $S_{+0}$ already
  integrated over $\Phi$ and evaluate (\ref{bound-1})
  to~(\ref{bound-5}).
\end{enumerate}
For bounds (\ref{bound-3}) and (\ref{bound-5}) the two procedures are
obviously equivalent, but for the other ones they are not.  While the
second possibility allows for a more inclusive measurement it is an
easy exercise to show that bounds (\ref{bound-1}), (\ref{bound-2}) and
(\ref{bound-4}) become weaker each time one integrates over a variable
before evaluating them, except if $S_{+0} /(S_{\veps} - S_{+-})$ is
constant in that variable---in this case procedures 1.\ and 2.\ give
again the same result. In practice this means that if $S_{\veps} -
S_{+-}$ and $S_{+0}$ have a quite different behaviour in a variable
$\Phi$ one can expect the bounds (\ref{bound-1}), (\ref{bound-2}),
(\ref{bound-4}) to be better if they are first evaluated with some
binning in $\Phi$ and then integrated. An example of such a variable
is the polar angle $\ta$ of the thrust axis defined in
Fig.~\ref{fig:tau} $(a)$, as we shall see.

We now show that there is a limit on how good the bounds
(\ref{bound-1}), (\ref{bound-2}), (\ref{bound-3}) can be. For this we
notice that there is an upper bound on the interference terms between
longitudinal and transverse photons:
\begin{equation}
  \label{upper-bound}
  2 \, |S_{+0}|^2 \le S_{00} \, (S_{++} - S_{+-})   \eqpt
\end{equation}
To see this it is convenient to change basis from circular to linear
photon polarisation vectors, related by $\veps_+ = - (\veps_1 + i
\veps_2) / \sqrt{2}$ and $\veps_- = (\veps_1 - i \veps_2) / \sqrt{2}$
where $\veps_1$ lies in the plane spanned by $\tau$ and $q$ in the
$\gamma^\ast p$ frame.  With the constrains from parity invariance
\cite{HDCB} one has $S_{10} = - \sqrt{2} \, S_{+0} \,$, $\, S_{11} =
S_{++} - S_{+-} \,$, $\, S_{22} = S_{++} + S_{+-}$ so that the above
inequality reads
\begin{equation}
  \label{cartesian-bound}
  |S_{10}|^2 \le S_{00} \, S_{11}  \eqpt
\end{equation}
Now we use that up to a flux and phase space factor $S_{m n}$ is given
by $\int d\Phi' \, {\cal A}_m^\ast {\cal A}_n^{\phantom{\ast}}$ where
${\cal A}_m$ is the amplitude of $\gamma^\ast p \to X \tilde{p}$ for
photon polarisation $m$ and $\Phi'$ denotes all variables over which
$S_{m n}$ is already integrated. We include in $\Phi'$ the
polarisations of $p$ and the final state particles, for which the
integral reduces to a sum. Taking the functions ${\cal A}_m(\Phi')$ as
elements of a Hilbert space and the integral over $\Phi'$ as a scalar
product (\ref{cartesian-bound}) is just the Schwarz inequality.

This argument also tells us that we have equality in
(\ref{upper-bound}), (\ref{cartesian-bound}) exactly if ${\cal A}_0$
and ${\cal A}_1$ are proportional to each other as functions of
$\Phi'$. In this case the l.h.s.\ of (\ref{bound-1}) and the r.h.s.\ 
of (\ref{bound-2}) reduce to $\frac{1}{2 \veps} (S_{++} - S_{+-} +
\veps S_{00}) \pm \frac{1}{2 \veps} | S_{++} - S_{+-} - \veps S_{00}
|$, i.e.\ to
\begin{equation}
  \label{best-bounds}
  S_{00} \hspace{3em} \mbox{and} \hspace{3em} \frac{S_{++} -
  S_{+-}}{\veps}  \eqcm
\end{equation}
so that one of the bounds on $S_{00}$ is $S_{00}$ itself, this can be
the lower or the upper bound. If (\ref{upper-bound}) is a strict
inequality then the bounds (\ref{bound-1}), (\ref{bound-2}) are less
good than in (\ref{best-bounds}). We see that they are rather far
apart if $S_{00}$ is much smaller or much bigger than $S_{++} - S_{+-}
= S_{11}$. For (\ref{bound-1}) and (\ref{bound-2}) to be tight bounds
one needs a region of phase space where the $\gamma^\ast p$ scattering
amplitudes with transverse and longitudinal photons are of comparable
magnitude and where they have a large enough interference.

As to the upper bound (\ref{bound-3}) one easily sees that it equals
$S_{00}$ if $S_{++} + S_{+-} = S_{22} = 0$ and is bigger otherwise.

\section{Particular diffractive final states}

\subsection{Two spin zero mesons}
To see that the above ``optimal bounds'' on $S_{00}$ can actually be
achieved in realistic cases let us consider a very simple diffractive
system $X = M \bar{M}$, where $M$ stands for a spin zero meson such as
a pion or kaon, and let $\tilde{p}$ be an elastically scattered
proton. In $S_{m n} = d \sigma_{m n} /(d \xp \, d \!\cos\ta)$ several
degrees of freedom $\Phi'$ are summed or integrated over:
\begin{enumerate}
\item the solid angle of the scattered proton in the $\gamma^\ast p$
  c.m. In the diffractive region it is a good approximation to replace
  this integration with taking the $\gamma^\ast p$ cross sections and
  interference terms at zero scattering angle and multiplying with a
  common overall factor. In the following we therefore consider the
  incoming and outgoing proton to be collinear in the $\gamma^\ast p$
  c.m.
\item the two configurations where the forward particle is $M$ or
  $\bar{M}$, related by swapping the meson momenta. Assuming that the
  $\gamma^\ast p$ reaction can be described by exchanges of positive
  charge conjugation parity between $p$ and $\gamma^\ast$, which of
  course holds if pomeron exchange dominates, it follows from charge
  conjugation invariance of the subreaction $\gamma^\ast +
  (\textrm{exchange}) \to M \bar{M}$ that the $\gamma^\ast p$ cross
  sections and interference terms are equal for these configurations,
  cf.~\cite{MD}.
\item the helicities $h$ and $\tilde{h}$ of initial and scattered
  proton in the $\gamma^\ast p$ frame; for zero scattering angle they
  are the same in the rest frame of $X$. We make the assumption that
  the $\gamma^\ast p$ amplitudes ${\cal A}_m^{h,\tilde{h}}$ satisfy
  ${\cal A}_m^{++} = {\cal A}_m^{--}$ and ${\cal A}_m^{+-} = {\cal
    A}_m^{-+} = 0$. This holds for instance in the limit of large
  $\gamma^\ast p$ c.m.\ energy in the two-gluon exchange model of
  Landshoff and Nachtmann~\cite{LN} and in the pomeron model of
  Donnachie and Landshoff~\cite{DL}.
\end{enumerate}
With these approximations the condition needed for (\ref{upper-bound})
to be an equality are satisfied and the bounds (\ref{bound-1}),
(\ref{bound-2}) take the form (\ref{best-bounds}). One can say more:
at $\ta = 0$ it follows from angular momentum conservation that the
$\gamma^\ast$ must be longitudinal and $S_{++} - S_{+-} = 0$. At $\ta
= \pi /2$ it is $S_{00}$ that must vanish: a rotation by $\pi$ about
the $z$ axis followed by charge conjugation of $\gamma^\ast +
(\textrm{exchange}) \to M \bar{M}$ gives ${\cal A}_0^{++} = - {\cal
  A}_0^{++}$ under our assumptions, the origin of the minus sign being
the negative charge conjugation parity of the photon. Assuming that
$S_{00}$ is not also zero at $\ta = 0$ or $S_{++} - S_{+-}$ at $\ta =
\pi /2$ we then have that $S_{00}$ in (\ref{best-bounds}) is the upper
bound for $\ta$ near 0 and the lower one for $\ta$ near $\pi /2$, and
at some value of $\ta$ the curves for the two bounds in
(\ref{best-bounds}) will cross over.

In the $X$ rest frame a parity transformation and subsequent rotation
by $\pi$ about the axis perpendicular to the scattering plane gives
${\cal A}_2^{++} = - {\cal A}_2^{--}$, where the minus sign comes from
the transformation of the photon polarisation $\veps_2$.  With ${\cal
  A}_2^{--} = {\cal A}_2^{++}$ we thus have $S_{++} + S_{+-} = S_{22}
= 0$ with our assumptions.\footnote{This result still holds if instead
  of being zero the amplitudes ${\cal A}_m^{+-}$ and ${\cal A}_m^{-+}$
  have equal size and an appropriate relative phase.} Hence bound
(\ref{bound-3}) is the longitudinal cross section itself.  For $\ta$
near $\pi /2$ where $S_{00}$ is smaller than $(S_{++} - S_{+-}) /
\veps = 2 S_{++} / \veps$ one has that both a lower (\ref{bound-1})
and an upper (\ref{bound-3}) bound are equal to $S_{00}$ which then is
completely constrained. Our assumptions in points 1.\ to 3.\ will of
course not be exactly satisfied but one can expect that very close
bounds on the longitudinal cross section can be obtained for $X = M
\bar{M}$ final states.

\subsection{$X = q \bar{q}$}

We now look at the diffractive final state at parton level, where
calculations have been made in several models of diffraction. The
simplest state is a quark-antiquark pair, for which detailed
predictions including the $\gamma^\ast p$ interference terms are
available in two-gluon exchange models~\cite{MD,BELW}. We will first
consider light quark flavours and neglect the quark mass.

If the quark and antiquark are only produced with opposite helicities,
which is the case for massless quarks in the two-gluon models cited,
and if one makes the same assumptions on the scattering of the proton
as in points 1.\ to 3.\ of the previous subsection, one finds again
that (\ref{upper-bound}) is an equality. Compared with $M \bar{M}$ one
now has an additional summation in $S_{m n} = d \sigma_{m n} /(d \xp
\, d \!\cos\ta)$ over the two $q \bar{q}$ helicity combinations.
Working in the c.m.\ of $X$ one can relate the corresponding
amplitudes by a parity transformation followed by rotation of $\pi$
about the axis perpendicular to the scattering plane and finds that
${\cal A}_0$ and ${\cal A}_1$ are again proportional as functions of
$\Phi'$.

Let us recall some results for the dependence of the $\gamma^\ast p$
cross sections and interference terms on $Q^2$ and on the transverse
momentum $P_T$ of the produced quark in the $\gamma^\ast p$ c.m.,
given by $\sin\ta = 2 P_T / M$ where $M$ is the invariant $q \bar{q}$
mass. If $P_T /M$ is small then $S_{00}$ and $S_{+-}$ are suppressed
by a factor $P_T^2 /M^2$ and $S_{+0}$ by a factor $P_T /M$ compared to
$S_{++}$ which dominates in this region, while at large $P_T$ all
terms can be of comparable magnitude. $S_{++}$ approximately falls
like $1 /P_T^4$ in the range $1 \GeV^2 \lsim P_T^2 \ll M^2$. The $P_T
\,$-integrated transverse cross section $d \sigma_{++} /d \xp$ is
dominated by small $P_T$ and behaves like $1 /Q^2$ at fixed $\xp$ and
$\beta$, which means Bjorken scaling of $F^D_T$. In contrast to this
the leading power is $1 / Q^3$ for $d \sigma_{+0} /d \xp$ and $1 /Q^4$
for $d \sigma_{00} /d \xp$ and $d \sigma_{+-} /d \xp$ so that in
particular $F^D_L$ is of higher twist. Note however that $F^D_T$
vanishes like $1 - \beta$ in the limit $\beta \to 1$ whereas $F^D_L$
is finite, so that in a region of sufficiently large $\beta$ and not
too large $Q^2$ the longitudinal structure function $F^D_L$ can be
appreciable. In such a region, where its role is most important, $c
\bar{c}$ production is suppressed or zero due to its production
threshold, which justifies the restriction of our discussion to light
flavours.  Independent of the quark mass one finds that $S_{+-}$ is
positive whereas $S_{+0}$ changes sign at some value of $\beta$ below
1/2, being positive below and negative above.

\begin{figure}
  \begin{center}  \leavevmode
    \setlength{\unitlength}{1cm}
    \begin{picture}(11.5,7.8)
      \put(0,0){\epsfxsize 10cm \epsfbox{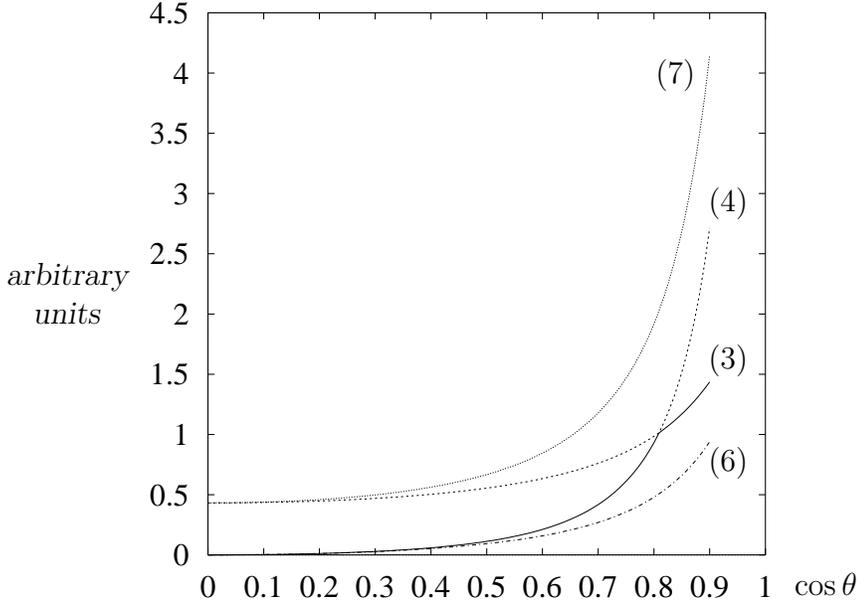}}
      \put(9.8,0.09){$\cos\ta$}
      \put(8.65,1.75){(\protect\ref{bound-4})}
      \put(8.65,3.1){(\protect\ref{bound-1})}
      \put(8.65,5.2){(\protect\ref{bound-2})}
      \put(7.95,6.9){(\protect\ref{bound-5})}
      \newlength{\dummy}
      \settowidth{\dummy}{\textsl{arbitrary}}
      \put(-0.7,4){\parbox{\dummy}{
          \begin{center}
            \textsl{arbitrary \\ units}
          \end{center}}}
    \end{picture}
  \end{center}
\caption{\label{fig:bounds}Bounds  (\protect\ref{bound-1}),
  (\protect\ref{bound-2}), (\protect\ref{bound-4}),
  (\protect\ref{bound-5}) on the longitudinal cross section $S_{00} =
  d \sigma_{00} / (d \xp \, d \!\cos\ta)$ for a final state $X = q
  \bar{q}$, calculated by two-gluon exchange~\protect\cite{MD}. The
  values of the relevant parameters are $Q^2 = 45 \GeV^2$, $\beta =
  0.9$, $\veps = 0.8$.}
\end{figure}

In Fig.~\ref{fig:bounds} we show an example of the $\ta$ dependence of
the differential bounds (\ref{bound-1}), (\ref{bound-2}) given by
(\ref{best-bounds}) and of their weaker versions (\ref{bound-4}),
(\ref{bound-5}). Since $S_{+-} \ge 0$ the bound (\ref{bound-3}) is not
useful in this case.  We see that the bounds (\ref{bound-1}) and
(\ref{bound-2}) are equal at some value of $\cos\ta$; at smaller
$\cos\ta$ the curve for $S_{00}$ coincides with the upper bound and at
larger $\cos\ta$ with the lower one. This crossover happens at $\cos^2
\ta \approx \frac{(2 \beta - 1)^2}{1 + 4 \, \beta (1 - \beta) (1
  /\veps - 1)}$ if the corresponding value of $P_T^2$ is above a few
GeV$^2$ so that certain approximations of $S_{m n}$ are valid. We also
find that bound (\ref{bound-4}) is rather close to (\ref{bound-1}) and
(\ref{bound-5}) to (\ref{bound-2}) at small $\cos\ta$, while at the
crossover point the ratio of (\ref{bound-4}) to (\ref{bound-1}) and of
(\ref{bound-2}) to (\ref{bound-5}) is easily found to be 1/2. We
remark that the lower bounds go to zero at $\cos\ta \to 0$ because for
any final state the interference term $S_{+0}$ vanishes at $\ta = \pi
/2$ due to symmetry reasons~\cite{MD}.  The curves in
Fig.~\ref{fig:bounds} stop at very large $\cos\ta$ where the
approximation used in their calculation becomes inaccurate. From an
experimental point of view it should be difficult to measure $\varphi$
if $\ta$ is below some critical value, this implies that an upper
bound can only be given for $F^D_L$ in a restricted kinematical
region, unless one is willing to extrapolate a measured upper bound on
$d \sigma_{00} /(d \xp \, d \!\cos\ta)$ down to $\ta = 0$.

It is worthwhile noting that the transverse-transverse interference
term $S_{+-}$ for $q \bar{q}$ is found to be positive, whereas for the
production of a $\pi^+ \pi^-$ pair we have seen in the previous
subsection that $S_{+-} = - S_{++} \le 0$. In other words the
preferred orientation of a quark-antiquark pair is perpendicular to
the electron plane in the $\gamma^\ast p$ c.m.\ while a pair of pions
prefers to be in that plane. Parton-hadron duality has recently been
invoked in~\cite{MartRysk} to calculate the production of $\pi \pi$
from $q \bar{q}$ in the region of low-lying resonances like the
$\rho$. If one takes this idea literally then the change of the
azimuthal distribution from $q \bar{q}$ to $\pi \pi$ is an interesting
effect of hadronisation---beyond the change in the $\ta$ distribution
imposed by angular momentum conservation.\footnote{At $\ta = 0$ we
  must have $S_{++} = 0$ for $\pi \pi$ but not for $q \bar{q}$ and
  $S_{00} = 0$ for $q \bar{q}$ but not for $\pi \pi$, assuming again
  that $q$ and $\bar{q}$ are produced with opposite helicities.} This
also implies that a parton level calculation for the angular
distribution cannot be used if the multiplicity of $X$ is too small.

\subsection{$X = q \bar{q} g$}

For the final state with a $q \bar{q}$ pair and an additional gluon no
complete calculation with two-gluon exchange has been performed yet.
Results in the leading $\alpha_s \log Q^2$ approximation have e.g.\ 
been reported in~\cite{Wuest}: the transverse structure function
$F^D_T$ for $q \bar{q} g$ behaves like $(1 - \beta)^3$ at large
$\beta$ and is negligible compared with the $q \bar{q}$ contribution
for $\beta > 1/2$, while $F^D_L$ for $q \bar{q} g$ is zero in this
approximation.

The three parton final state has also been investigated in the
semiclassical model of~\cite{Buch} who find that it gives leading
twist contributions both to $F^D_T$ and $F^D_L$. In \cite{Heb} it was
shown that this approach can be reformulated in terms of the
diffractive parton model: the proton emits a parton which scatters on
the $\gamma^\ast$, producing two of the partons in $X$. It is required
that their transverse momentum in the $\gamma^\ast p$ frame be
sufficiently large for this scattering to be hard. The third parton in
$X$ is approximately collinear with the proton and plays the role of a
``pomeron remnant''.

Let us then take a closer look at what the parton model description
gives for the longitudinal cross section and for the $\gamma^\ast p$
interference terms with the final states just described. The
calculation is completely analogous to the one for the azimuthal
dependence in nondiffractive deep inelastic scattering with two
partons and a proton remnant in the final state, which can e.g.\ be
found in~\cite{Oldies}.

Call the four-momenta of the two partons produced in the hard
scattering $P_1$ and $P_2$, with $P_1$ being the forward particle,
i.e.\ having the larger longitudinal momentum along the photon
direction in the c.m.\ of $X$. Let further be $P_T$ the transverse
momentum of $P_1$ in that frame and $\hat{s} = (P_1 + P_2)^2$. We
first give $\gamma^\ast p$ interference terms defined with respect to
the azimuthal angle $\varphi'$ between the electron momentum $k$ and
the vector $\tau' = (P_1 - P_2) / \hat{s}$. Restricting our analysis
to sufficiently large $P_T$ the effects of a nonzero transverse
momentum of the parton emitted by the proton (and thus of the pomeron
remnant) should not be too large and we set this transverse momentum
to zero.\footnote{Intrinsic transverse parton momentum in
  nondiffractive $ep$ scattering has been investigated
  in~\cite{Transverse}.} We then have
\begin{eqnarray}
  \label{proton-parton}
  \frac{d \sigma_{m n}}{d \xp} &=& \sum_q \frac{\pi  \alpha_s
    \alpha_{\it em} e_q^2}{(1 - x) \, Q^2} \, \int_{\beta}^{\bh_{\it
    max}} d \bh \, \int_{P_{\it min}^2}^{\hat{s}/4} \frac{d P_T^2}{
    \sqrt{\hat{s} (\hat{s}/4 - P_T^2)} }  \cdot  \\ \nonumber
  && \frac{\beta}{\bh} \cdot \left[ g\left( \frac{\beta}{\bh}, \xp
    \right) T_{m n}^{q \bar{q}} + q\left( \frac{\beta}{\bh}, \xp
    \right) T_{m n}^{g q} + \bar{q}\left( \frac{\beta}{\bh},
    \xp \right) T_{m n}^{g q} \right]  \eqcm
\end{eqnarray}
where $\bh = Q^2 /(Q^2 + \hat{s})$ and its upper limit $\bh_{\it max}$
follows from the lower cutoff on $P_T$. The sum $\sum_q$ is over the
flavours of the quarks and $e_q$ denotes their charge in units of the
positron charge.  For simplicity we have taken the quarks to be
massless, neglecting the complications for charm production. $g(z,
\xp)$, $q(z, \xp)$ and $\bar{q}(z, \xp)$ respectively are the
diffractive gluon, quark and antiquark distributions for a momentum
fraction $z$ of the parton with respect to the momentum transfer
$\Delta$ from the proton. They are integrated over $t = \Delta^2$, so
that to leading order in $\alpha_s$ one has $F^{D(3)}_2(\xp, \beta,
Q^2) = \sum_q e_q^2 \, \beta \, \Big( q(\beta,\xp) +
\bar{q}(\beta,\xp) \Big)$.  Finally we have
\begin{eqnarray}
  \label{BGF}
  T_{++}^{q \bar{q}} & = & \frac{1}{2} \cdot 8 \, \Big( 1 - 2 \bh
  (1 - \bh) \Big) \left[ \frac{\hat{s}/4 - P_T^2}{P_T^2} +
    \frac{1}{2} \right]  \\ \nonumber
  T_{00}^{q \bar{q}} & = & \frac{1}{2} \cdot 16 \, \bh (1 - \bh)
  \\ \nonumber
  T_{+-}^{q \bar{q}} & = & - \frac{T_{00}^{q \bar{q}}}{2}
  \\ \nonumber
  T_{+0}^{q \bar{q}} & = & \frac{1}{2} \cdot \frac{8}{\sqrt{2}} \,
    \frac{\sqrt{\hat{s}/4 - P_T^2}}{P_T} \, \sqrt{\bh (1 - \bh)} \, (2
  \, \bh - 1)
\end{eqnarray}
for boson-gluon fusion $\gamma^\ast g \to q \bar{q}$ and
\begin{eqnarray}
  \label{QCD-C}
  T_{++}^{g q} & = & \frac{4}{3} \cdot \frac{1}{1 - \bh}
    \left[ 4 \, (1 + \bh^2) \, \frac{\hat{s}/4 - P_T^2}{P_T^2} + 5 - 2
    \, \bh (1 - \bh) \right]  \\ \nonumber
  T_{00}^{g q} & = & \frac{4}{3} \cdot 4 \, \bh  
  \\ \nonumber 
  T_{+-}^{g q} & = & - \frac{T_{00}^{g q}}{2}
  \\ \nonumber
  T_{+0}^{g q} & = & \frac{4}{3} \cdot \frac{4}{\sqrt{2}} \,
    \frac{\sqrt{\hat{s}/4 - P_T^2}}{P_T} \,
    \frac{\sqrt{\bh^3_{\phantom{0}}}}{\sqrt{1 - \bh}}
\end{eqnarray}
for the QCD Compton processes $\gamma^\ast q \to g q$ and $\gamma^\ast
\bar{q} \to g \bar{q}$. We see that at small $P_T^2 / \hat{s}$ the
$\gamma^\ast p$ cross sections and interference terms have the same
relative factors of $1 /P_T$ as in the case $X = q \bar{q}$ so that in
this region the transverse cross section dominates. The absolute
behaviour in $P_T$ is however different; integrating over $P_T$ one
finds that $d \sigma_{00} / d\xp$, $d \sigma_{+-} / d\xp$ and $d
\sigma_{+0} / d\xp$ behave like $1 /Q^2$ at fixed $\beta$ and $\xp$,
corresponding to leading twist contributions to the $ep$ cross
section, whereas $d \sigma_{++} / d\xp$ has a collinear singularity at
$P_T = 0$ which with an appropriate cutoff gives a leading twist
contribution enhanced by $\log Q^2$, as it is also found in the
two-gluon exchange calculation~\cite{Wuest}.

Looking at the region of large $\beta$ we see in (\ref{BGF}),
(\ref{QCD-C}) that the longitudinal cross section is suppressed
compared with the transverse one by a factor $(1 - \bh) \le (1 -
\beta)$, both for boson-gluon fusion and QCD Compton scattering. The
behaviour of $F^D_L$ in the large-$\beta$ limit depends on how the
diffractive parton distributions behave for $z \to 1$. If one assumes
a power behaviour $g(z, \xp) \sim (1 - z)^{n_g}$ and $q(z, \xp), \,
\bar{q}(z, \xp) \sim (1 - z)^{n_q}$ with exponents $n_g, n_q > -1$
then $F^D_L$ is bounded from above by $c_g \, (1 - \beta)^{n_g + 2} +
c_q \, (1 - \beta)^{n_q + 1}$ with some constants $c_g$, $c_q$.  It
was argued in~\cite{BS} that the behaviour of the parton distributions
should be between $(1-z)^0$ and $(1-z)^1$ for gluons and between
$(1-z)^1$ and $(1-z)^2$ for quarks; in this case $F_L^D$ would vanish
at least like $(1 - \beta)^2$.  In~\cite{Kwiec} the ratio $F^D_L
/F^D_T$ was calculated in the diffractive parton model with a
particular ansatz for the parton distributions and indeed came out
small for $\beta \ge 1/2$; it is on the contrary at small $\beta$
where this ratio was found to be appreciable.

Let us now investigate the bounds one can obtain for $S_{00}$, first
for the differential quantities $S_{m n} = d \sigma_{m n} /(d \xp\, d
P_T^2 \, d \bh)$. From (\ref{proton-parton}) to (\ref{QCD-C}) one can
show that for all values of the kinematic variables the expansion of
the square roots in (\ref{bound-1}), (\ref{bound-2}) which leads to
the simplified bounds (\ref{bound-4}), (\ref{bound-5}) is an
approximation better than 5\% so that it is enough to discuss the
latter. Unlike in the case $X = q \bar{q}$ one now finds that
(\ref{upper-bound}) is always a strict inequality; in fact already the
summation over particle helicities in the diffractive final state
violates the conditions for equality in (\ref{upper-bound}). It turns
out that now $4 \, |S_{+0}|^2 \le S_{00} \, (S_{++} - S_{+-})$ with a
factor $4$ instead of $2$ on the l.h.s. The ratio of right and left
hand side goes to 1 if $P_T \to 0$ and $\bh = 1$ for QCD Compton
scattering and if $P_T \to 0$ and $\bh = 1$ or $\bh = 0$ for
boson-gluon fusion. With this we have that the lower bound
(\ref{bound-4}) is at most $0.5 \cdot S_{00}$. To give a numerical
example away from the edges of phase space we take $4 P_T^2 / \hat{s}
\ge 0.2$, $0.1 \le \bh \le 0.9$ and $\veps = 0.8$ and find that the
bound is between 0 and $0.33 \cdot S_{00}$ for boson-gluon fusion and
between 0 and $0.38 \cdot S_{00}$ for QCD Compton scattering.

The upper bound (\ref{bound-3}) is now better than (\ref{bound-5})
because $S_{+-} \le 0$, and becomes good where $S_{++} + S_{+-}$ is
not large compared to $S_{00}$. From (\ref{BGF}), (\ref{QCD-C}) we see
that this is only the case if $4 P_T^2 /\hat{s}$ is large enough.
Comparing $T_{++}^{q \bar{q}} + T_{+-}^{q \bar{q}} \ge \frac{1}{2}
\cdot 4 \, (2 \bh - 1)^2$ with $T_{00}^{q \bar{q}}$ and $T_{++}^{g q}
+ T_{+-}^{g q} \ge \frac{4}{3} \cdot (1 - \bh)^{-1}$ with $T_{00}^{g
  q}$ we further see that $\bh (1 - \bh)$ must not be small.  For
$\veps = 0.8$, $0.2 \le \bh \le 0.8$ and $4 P_T^2 / \hat{s} = 0.5$ we
find an upper bound between $2.2 \cdot S_{00}$ and $4.4 \cdot S_{00}$
for boson-gluon fusion and between $12 \cdot S_{00}$ and $22.5 \cdot
S_{00}$ for the QCD Compton process, when going down to $4 P_T^2 /
\hat{s} = 0.1$ these bounds become about five times larger.

We now have to see how the interference terms corresponding to the
vector $\tau$ defined from the thrust axis are related to those
discussed so far. We recall that for a system $X = q \bar{q} g$ with
zero quark mass the thrust axis in its rest frame is given by the
direction of the most energetic particle. This can be (\textit{i}) the
forward parton produced in the hard $\gamma^\ast$ parton collision or
(\textit{ii}) the parton playing the role of a pomeron remnant. For
events of type (\textit{i}) we have $\varphi = \varphi'$, i.e.\ $\tau$
and $\tau'$ lead to the same $\gamma^\ast p$ interference terms given
in (\ref{proton-parton}) to (\ref{QCD-C}).  In our simple calculation
with zero transverse momentum for the pomeron remnant events of type
(\textit{ii}) have $\tau$ collinear with $q$ and $p$ and do not
contribute to the $\varphi$-asymmetry in the $ep$ cross sections, the
corresponding interference terms thus are zero~\cite{MD}. The
condition for (\textit{i}) is $1 - \sqrt{1 - 4 P_T^2 / \hat{s}} < 2
\beta (1 - \bh) / (\bh - \beta)$. It is always fulfilled for $\bh < 3
\beta / (2 \beta + 1)$ and otherwise only for $P_T$ below some
critical value.

Unless one attempts a separation of final states $q \bar{q}$ and $q
\bar{q} g$, using for instance the value of the thrust, one will sum
over them when evaluating the $S_{m n}$. To investigate their relative
importance is beyond the scope of our study, but our arguments have
shown that in regions of phase space where $q \bar{q} g$ states
dominate the interference terms and the longitudinal cross section the
bounds will not be very tight, whereas quite good bounds can be
expected where the $q \bar{q}$ state dominates.

Beyond the possibility to obtain bounds on $F_L^D$ the $\gamma^\ast p$
interference terms are interesting in themselves. From
(\ref{proton-parton}) and (\ref{BGF}), (\ref{QCD-C}) we see that both
for boson-gluon fusion and QCD Compton scattering the
transverse-transverse interference is negative and thus has the
opposite sign than what we found for $X = q \bar{q}$ (cf.\ 
also~\cite{BELW}). The transverse-longitudinal interference is more
complicated, but if $\beta > 1/2$ one has $2 \bh - 1 > 0$ and it is
positive for $X = q \bar{q} g$ in our parton model calculation and
thus again opposite to the one for $X = q \bar{q}$ calculated by
two-gluon exchange. In this sense the sign of the interference terms
gives a hint on the underlying final state and its production
mechanism.

Another difference between the final states is the leading power
behaviour in $1 /Q$ at fixed $\beta$ and $\xp$ of the integrated
interference terms and cross sections $S_{m n} = d \sigma_{m n} / d
\xp \,$. For $X = q \bar{q}$ we found $S_{+-} \sim 1 /Q^4$ and $S_{+0}
\sim 1 /Q^3$ compared with $ S_{\veps} \sim 1 /Q^2$ whereas for $X = q
\bar{q} g$ all three terms go like $1 /Q^2$. There will be logarithmic
corrections to these powers, but unless they strongly differ for
$S_{+-}$, $S_{+0}$ and $S_{\veps}$ the relative behaviour of the
interference terms and the sum of cross sections is clearly distinct
in the two cases. More generally the inequality (\ref{upper-bound})
connects the $Q^2$ dependence of $S_{00}$ and $S_{+0} \,$: an
interference $S_{+0}$ that only decreases like $1 /Q^2$ excludes a
nonleading leading twist behaviour of $S_{00}$ beyond some value of
$Q^2$ if we assume that $S_{++} - S_{+-}$ is leading twist (experiment
indicates that $S_{\veps}$ is).

\section{Conclusions}

The distribution of an azimuthal angle defined with the help of the
thrust axis of the diffractive final state allows to extract
interference terms between different polarisations of the exchanged
photon in diffractive deep inelastic scattering. They may help to
answer the important question of whether the cross section for
longitudinal photons is leading twist or not and furthermore give
information on which diffractive final states dominate in a given
kinematic region.

These interference terms can be used to obtain model independent
bounds on the longitudinal cross section. We have shown that it can be
advantageous to evaluate these bounds first with some additional
binning in variables like the polar angle $\ta$ of the thrust axis.
Such differential bounds can be equal to the longitudinal cross
section itself. For diffractive final states $\pi \pi$ or $K K$  this
happens under weak dynamical assumptions, which should be
satisfied to a good approximation in the diffractive regime.

Using the results of two-gluon exchange models one has that the $q
\bar{q}$ diffractive final state gives a longitudinal contribution
$F^D_L$ to $F^D_2$ which is suppressed by $1 /Q^2$ but can be
non-negligible at large $\beta$. The estimated bounds one could obtain
on $F^D_L$ in this kinematic region look quite good, especially if one
evaluates them first binned in $\ta$. To investigate $q \bar{q} g$
final states we used the diffractive parton model. We have shown that
one does not expect these final states to lead to an appreciable ratio
$F^D_L /F^D_T$ at large $\beta$, but remark that a ratio of up to 0.5
was found at small $\beta$ in~\cite{Kwiec}. In an estimation
neglecting the effects of intrinsic parton momentum and hadronisation,
which should be valid if there is large enough $P_T$ in the
diffractive system we find that if this final state dominates then the
bounds on $F_L^D$ obtained from the interference terms are much less
stringent than in the $q \bar{q}$ case, except in some corners of
phase space.

\section*{Acknowledgments}
I gratefully acknowledge conversations with J. Bartels, A. Hebecker,
M. McDermott and M. W\"usthoff. CPT is Unit\'e Propre 14 du Centre
National de la Recherche Scientifique.


\begin{thebibliography}{99}

\newcommand{\journal}[4]{{#1} #2 (#3) #4}
\newcommand{\NPB}{Nucl.\ Phys.\ B}
\newcommand{\PLB}{Phys.\ Lett.\ B}
\newcommand{\PRL}{Phys.\ Rev.\ Lett.\ }
\newcommand{\PRD}{Phys.\ Rev.\ D}
\newcommand{\ZPC}{Z.\ Phys.\ C}
\newcommand{\JPG}{J.\ Phys.\ G}

\bibitem{HERA} H1 Collaboration, ``Inclusive Measurement of
  Diffractive Deep-inelastic $ep$ Scattering'', paper submitted to the
  International Europhysics Conference on High Energy Physics, HEP97,
  Jerusalem, Israel, August 1997, preprint DESY--97--158
  
\bibitem{Brisk} G. Briskin and M. F. McDermott, ``Diffractive
  structure functions in DIS'', in: \emph{Future Physics at HERA},
  Proc.\ of the Workshop 1995/96, eds.\ G. Ingelman et al., DESY 1996
  
\bibitem{BarTwist} J. Bartels, talk given at the Madrid Workshop on
  low-$x$ Physics, Miraflores de la Sierra, Spain, 18--21 June 1997,
  to appear in the proceedings

\bibitem{HDCB} T. Arens, O. Nachtmann, M. Diehl and P. V. Landshoff,
  \journal{\ZPC}{74}{1997}{651}

\bibitem{MD} M. Diehl, hep-ph/9610430
  
\bibitem{LN} P. V. Landshoff and O. Nachtmann,
  \journal{\ZPC}{35}{1987}{405}

\bibitem{DL} A. Donnachie and P. V. Landshoff,
  \journal{\NPB}{244}{1984} 322

\bibitem{BELW} J. Bartels, C. Ewerz, H. Lotter and M. W\"usthoff,
  \journal{\PLB}{386}{1996}{389}
  
\bibitem{MartRysk} A. D. Martin, M. G. Ryskin, and T. Teubner,
  \journal{\PRD}{55}{1997}{4329}; \journal{\PRD}{56}{1997}{3007}

\bibitem{Wuest} M. W\"usthoff, hep-ph/9702201 

\bibitem{Buch} W. Buchm\"uller, M. F. McDermott and A. Hebecker,
  \journal{\NPB}{487}{1997}{283}

\bibitem{Heb} A. Hebecker, hep-ph/9702373
  
\bibitem{Oldies} H. Georgi and J. Sheiman,
  \journal{\PRD}{20}{1979}{111}

\bibitem{Transverse} A. K\"onig and P. Kroll,
  \journal{\ZPC}{16}{1982}{89}; \\
  A. S. Joshipura and G. Kramer, \journal{\JPG}{8}{1982}{209}
  
\bibitem{BS} A. Berera and D. E. Soper, \journal{\PRD}{53}{1996}{6162}

\bibitem{Kwiec} K. Golec-Biernat and J. Kwieci\'nski,
  \journal{\PLB}{353}{1995}{329}
\end{thebibliography}
\end{document}